\documentstyle[epsfig,12pt]{article}
\newcommand{\bc}{\begin{center}}
\newcommand{\ec}{\end{center}}
\newcommand{\bt}{\begin{tabular}}
\newcommand{\et}{\end{tabular}}

\begin{document}
\bc
{\Large\bf
Okubo-Zweig-Iizuka Rule Violation and Possible Explanations}\footnote
{Invited talk for the Third International Conference on Nucleon-Antinucleon
Physics (NAN'95), Moscow, Sept. 11-16, 1995}\\[1.5cm]

Bing-song ZOU

Queen Mary and Westfield College, London E1 4NS, United Kingdom\\[1cm]

{\bf Abstract}
\ec

Violations of the Okubo-Zweig-Iizuka (OZI) rule in $s\bar s \leftrightarrow 
n\bar n\!\equiv\!\frac{1}{\sqrt{2}}(u\bar u + d\bar d)$ mixing and $\phi$ 
production from $N\bar N$ annihilation are reviewed. Possible explanations 
are examined. We conclude that the two-step hadronic loops can explain
these OZI violations naturally with proper consideration of cancellations
between loops. No conclusive evidence exists for glueball states,
$s\bar sn\bar n$ four quark states, instanton effects, and strange quarks 
in nucleon.

\vskip 1cm
\section{ Introduction}

The empirical OZI rule [1] has been proposed thirty years ago without a
solid theoretical basis. Its usual statement is that diagrams with 
disconnected quark lines are negligible compared to those with connected 
quark lines, though there are other definitions in the literature [2]. 
A typical example is shown in Fig.1, where Fig.1b should be negligible 
compared with Fig.1a according to the rule. The narrow decay widths 
($<0.1$ MeV) of $J/\Psi$ and $\Upsilon(1S)$ states are clear evidence 
supporting the OZI rule, since these decays involve the annihilation of 
the $c\bar c$ or $b\bar b$ quarks corresponding to Fig.1b. 
As a comparison the decay widths of $\Psi(4040)$ and
$\Upsilon(10860)$ are larger than 50 MeV, corresponding to Fig.1a.

The most extensive experimental tests of the OZI rule were on $\phi$ and
$f'_2$ production; $\phi$ and $f'_2$ are close to pure $s\bar s$ states.
With the assumption that the coupling of the $\phi$ and $f'_2$ to nonstrange
hadrons are entirely due to their small nonstrange $n\bar n$ admixture parts,
a stronger version of the OZI rule, named the ``universal mixing model",
predicts [3]
\begin{equation}
\frac{\sigma(\pi N\to\phi X)}{\sigma(\pi N\to\omega X)}
=\frac{\sigma(N\bar N\to\phi X)}{\sigma(N\bar N\to\omega X)}
=tan^2\delta_V
\end{equation}
\begin{equation}
\frac{\sigma(\pi N\to f'_2 X)}{\sigma(\pi N\to f_2 X)}
=\frac{\sigma(N\bar N\to f'_2 X)}{\sigma(N\bar N\to f_2 X)}
=tan^2\delta_T
\end{equation}
where $X$ denotes any single or multiparticle final state containing
no strange particles; the $\delta_V$ and $\delta_T$ are $s\bar s$ - $n\bar n$
mixing angles for vector and tensor mesons, respectively. 
Usually when we talk about OZI rule violation,
it means violation of this stronger version of the OZI rule.

Okubo [4] has reviewed the experimental evidence for the OZI rule, with
emphasis on pion-induced reactions and meson decays. He concluded that
the experimental data on $\phi$ and $f'_2$ at that time 
(1977) were reasonably consistent with the validity of the rule. 
In spite of its reasonable successes, the simple formulation of the
OZI rule suffers an intrinsic logical flaw since OZI-forbidden
processes can take place as the product of two OZI-allowed processes.
This is the so-called ``higher-order paradox" of Lipkin [5]. Lipkin clarified
the microscopic origins of the OZI rule by showing how cancellations 
occur between the contributions of various hadronic loops. For example,
for $s\bar s$-$n\bar n$ mixing, the $K^*\bar K$ and $\bar K^* K$ loops
always have opposite phase to $K\bar K$ and $K^*\bar K^*$ loops.
This sort of cancellation was also showed by T\"ornqvist's unitarity quark
model [6] and Geiger-Isgur's calculations of hadronic loop contributions
to meson propagators [7]. The degree of cancellation varies for different
$q\bar q$ nonets [6,7].

Recently, abundant $\phi$-meson production in $N\bar N$ annihilation was
observed by ASTERIX, CRYSTAL BARREL, JETSET and OBELIX collaborations at LEAR.
Several $\phi$ production channels have branching ratios more than one
order of magnitude larger than predictions of the OZI universal mixing model.
The substantial OZI rule violations are intriguing and were described as
evidence for glueball states [8], $s\bar sn\bar n$ four quark states [9],
instanton effects [10] and the considerable admixture of $s\bar s$ 
components in the nucleon [11,12]. 
However it was shown [13-20] that the conventional
hadronic loop diagrams can also explain these large enhancements.

In this talk, first I will show for $s\bar s$-$n\bar n$ mixing how the
OZI rule evades large hadronic loop corrections for some $q\bar q$ nonets
but is scuttled for other nonets. Secondly, I will review the hadronic loop
contributions to $\phi$ production from $N\bar N$ annihilation.
Then I will briefly introduce other possible explanations and examine whether 
there are any clear-cut predictions to distinguish them from the 
conventional hadronic loop
mechanism in $N\bar N$ annihilations. Finally I will give my conclusion.

\section{OZI rule and $s\bar s$-$n\bar n$ mixing}

For the $s\bar s$-$n\bar n$ mixing, the simplest quark line diagram
shown by Fig.2a is an OZI forbidden process. If hadronic loop diagrams
were negligible, the OZI rule would predict very small mixing angles. 
In Table 1 we list the mixing angles obtained from experimental data [6,21] 
for the low-lying $q\bar q$ nonets. Only for the $1^{--}$ nonet is the mixing 
angle close to zero. The mixing angles are still
reasonably small for $2^{++}$ and $3^{--}$ nonets, but quite large 
for other nonets. So the OZI rule seems to be not working very well here.
A natural explanation for this is that the hadronic loop diagrams shown
by Fig.2b are not negligible. In fact the imaginary part of the loop
amplitudes are fixed to be non-zero by the unitarity relation:
\begin{equation}
Im T_{s\bar s\to n\bar n}=\sum_c\rho_cT^\dagger_{s\bar s\to c}T_{c\to n\bar n} .
\end{equation}
Here $c$ is a common channel for $s\bar s$ and $n\bar n$ decays, such as
$K\bar K$ or $K^*\bar K$; $\rho_c$ is the phase space factor for channel $c$.

A simple estimation of the hadronic loop contributions can be made
by considering the mass matrix in the basis of {$s\bar s$, $n\bar n$}:
\begin{equation}
\hat M=\pmatrix {m_{s\bar s}-\frac{i}{2}\Gamma_{s\bar s} &
\sum_c(A_c-\frac{i}{2}\epsilon_c\sqrt{\Gamma_{s\bar s\to c}
\Gamma_{n\bar n\to c}}) \cr
\sum_c(A_c-\frac{i}{2}\epsilon_c\sqrt{\Gamma_{s\bar s\to c}
\Gamma_{n\bar n\to c}}) &
m_{n\bar n}-\frac{i}{2}\Gamma_{n\bar n} } .
\end{equation}
Here $\epsilon_c=\pm 1$ is the relative phase for loop $c$. Neglecting
loop diagrams is equivalent to assuming a diagonal real mass matrix.
In other words, all the imaginary parts and the real off-diagonal parts 
($A_c$) of the mass matrix are coming from loop diagrams. All the
imaginary parts of the mass matrix come from on-shell loops and their values 
at the mass of a corresponding resonance 
are well determined by the partial decay widths measured by experiments
except relative phases $\epsilon_c$. The real off-diagonal parts $A_c$
come from virtual off-shell loops and can be obtained by dispersive relation 
from the energy dependent
imaginary parts or from some quark model calculations [7]. But they are
very model dependent. The physically observed states should be eigenstates
of the mass matrix and therefore must be $s\bar s$-$n\bar n$ mixed states.
Generally speaking, the larger the off-diagonal parts are, the bigger
the $s\bar s$-$n\bar n$ mixing will be.
If the off-diagonal parts are much smaller than $(m_{s\bar s}-m_{n\bar n})$,
then the mixing angle can be obtained perturbatively:
\begin{equation}
|sin\delta|^2\approx \frac{(\sum_c A_c)^2+\frac{1}{4}
\left[\sum_c\epsilon_c\sqrt{\Gamma_{s\bar s\to c}(s)
\Gamma_{n\bar n\to c}(s)}\right]^2}{(m_{s\bar s}-m_{n\bar n})^2} .
\end{equation}

The unitarity limit is obtained by assuming the $A_c$ part to be zero and gives
a lower limit for the mixing. For example, for $1^{--}$ and $2^{++}$ nonets,
at $\phi$ and $f'_2(1525)$ masses, the only observed on-shell strange meson 
loop is $K\bar K$. From quark flavor SU(3) symmetry, 
we have $\Gamma_{n\bar n\to K\bar K}=\Gamma_{s\bar s\to K\bar K}/2$.
Then in the unitarity limit their $s\bar s$-$n\bar n$ mixing angles are 
given by
\begin{equation}
|sin\delta|\approx\frac{\Gamma_{s\bar s\to K\bar K}}
{2\sqrt{2}(m_{s\bar s}-m_{n\bar n})} .
\end{equation}
Approximating $\phi$ - $f'_2(1525)$ as $s\bar s$ and $\omega$-$f_2(1270)$
as $n\bar n$, using PDG [21] mass and width values, the above equation
gives $\delta_V=0.3^\circ$ and $\delta_T=4.9^\circ$. As lower limits, they 
are compatible with the observed values listed in Table~1. The very small
contribution from the on-shell $K\bar K$ loop for $\phi$ and $f'_2(1525)$ is
due to very small $K\bar K$ phase space for $\phi$ and suppression by the
centrifugal barrier factor for $l>0$ decays. The puzzle is why the 
contribution from virtual loops, $A_c$, should also be small for the 
$1^{--}$ and $2^{++}$ nonets, as implied by their observed mixing angles.     
Lipkin suggested an explanation. He gave a general deduction [5] that 
$K\bar K^*$ and $\bar KK^*$ loops have opposite phase to $K\bar K$ and
$K^*\bar K^*$ loops for $s\bar s$-$n\bar n$ mixing. It is these loop
cancellations that make $A_c$ very small for $1^{--}$, $2^{++}$ and $3^{--}$ 
nonets.  This sort of cancellation was also shown by T\"ornqvist's 
unitarized quark model [6] and Geiger-Isgur's calculations of hadronic loop
contributions to meson propagators [7].

\noindent Table 1. $s\bar s$-$n\bar n$ mixing angles $\delta$ [6,21] and
corresponding hadronic loop contributions. $A'(m_{s\bar s})$ are mixing 
amplitudes from the $^3P_0$ model [7] in units of MeV. 
$+$ and $-$ present the relative phase of loops. $0$ stands for
forbidden. 
\begin{center} 
\begin{tabular}{|l|c|c|c|c|c|c|c|}
\hline
$J^{PC}$ & $|\delta(m_{s\bar s})|$ & $A'(m_{s\bar s})$ 
& $K\bar K$ & $K\bar{K^*}$ & $K^*\bar K$ & $K^*\bar K^*$ \\\hline
$1^{--}$ & $0.7\sim 3.4^\circ$ [21] & 8.6 & $+$ & $-$ & $-$ & $+$ \\
$2^{++}$ & $7\sim 9^\circ$ [21] & -7.1 & $+$ & $-$ & $-$ & $+$ \\
$3^{--}$ & $6\sim 7^\circ$ [21] & 7.4 & $+$ & $-$ & $-$ & $+$ \\
$1^{+-}$ & $\sim 18^\circ$ [6] & -59.4 & $0$ & $-$ & $-$ & $+$ \\
$1^{++}$ & $\sim 26^\circ$ [6] & 89.5 & $0$ & $-$ & $-$ & $+$ \\
$0^{-+}$ & $45\sim 58^\circ$ [21] & & $0$ & $-$ & $-$ & $+$   \\
$0^{++}$ & $\sim 36^\circ$ [6] & -537 & $+$ & $0$ & $0$ & $+$  \\\hline
\end{tabular}
\end{center}

Now the question is why the hadronic loop contribution is much larger for
other nonets shown in Table 1. Geiger and Isgur suggested [7] that it is
due to $^3P_0$ dominance of the effective quark-antiquark pair creation
operator which gives different $s\bar s\leftrightarrow n\bar n$ amplitudes 
for different nonets. Using the $^3P_0$ model, they calculated the real 
parts of $s\bar s$-$n\bar n$
mixing amplitudes, $A'(m_{s\bar s})$, which are listed in Table 1 and
consistent with the observed mixing angles. However we do not think
that the $^3P_0$ mechanism is the main reason and have suggested a more
general model-independent explanation [22], i.e., for some nonets either
$K\bar K$ or $K^*\bar K+K\bar K^*$ loops are forbidden by parity
conservation. In Table 1, we list the relative
phase from each hadronic loop for the low-lying nonets, while 0 stands
for forbidden. For $1^{--}$, $2^{++}$ and $3^{--}$ nonets, all four
loops are allowed and we expect the largest cancellations; for
$1^{+-}$, $1^{++}$ and $0^{-+}$ nonets, the $K\bar K$ loop is forbidden
and we expect weaker cancellations; for the $0^{++}$ nonet,  
$K^*\bar K + K\bar K^*$  loops are forbidden and there is {\bf NO} 
cancellation!  These simple model independent expectations are 
consistent with both observed mixing angles and the $^3P_0$ model
calculations [7]. For $0^{-+}$ nonet, the large
mixing angle can also be explained by hadronic loops [6] though 
its U(1) anomaly explanation is not excluded.

Due to the large $s\bar s$-$n\bar n$ mixing for $0^{-+}$ nonets,
$\eta\eta$, $\eta\eta'$ and $\eta'\eta'$ loops can also contribute to 
the $s\bar s$-$n\bar n$ mixing of some nonets. However, according to
the flavor SU(3) symmetry [23]:
\begin{eqnarray*}
<f'|\eta\eta><\eta\eta|f>&=&\frac{sin^2(2\delta_P)}{8}<f'|K\bar K>
<K\bar K|f> ,\\
<f'|\eta'\eta'><\eta'\eta'|f>&=&\frac{sin^2(2\delta_P)}{8}<f'|K\bar K>
<K\bar K|f> ,\\
<f'|\eta\eta'><\eta\eta'|f>&=&-\frac{sin^2(2\delta_P)}{4}<f'|K\bar K>
<K\bar K|f> .
\end{eqnarray*}
So the contribution from $\eta\eta$, $\eta\eta'$, $\eta'\eta'$ loops is a 
second order effect and much smaller than strange meson loops 
for $s\bar s$-$n\bar n$ mixing of $J^{PC}=(even)^{++}$ nonets. Note also that
the $\eta\eta$ loop has the same phase as $K\bar K$. The $\eta\eta$ loop
is forbidden for $0^{-+}$, $1^{--}$, $1^{+-}$, $1^{++}$ and $3^{--}$ nonets.

There is another important point for the $0^{++}$ nonet. The on-shell 
$K\bar K$ loop can give a very large imaginary part to the 
$s\bar s\leftrightarrow n\bar n$ transition amplitude because no
centrifugal barrier factor is present here for S-wave decay. The large
coupling to $K\bar K$ is also
the reason for the narrow peak structure of $f_0(980)$ [6,24].
Due to the very large $K\bar K$ loop contribution and no cancellations, 
there should not exist nearly pure $s\bar s$ $0^{++}$ mesons. 

In summary, the hadronic loop mechanism can explain naturally the 
$s\bar s$-$n\bar n$ mixing for all low-lying nonets with a proper
consideration of loop cancellations. Therefore for other OZI violations,
we should first examine the hadronic loop contributions.

\section{Hadronic loop mechanism for the abundant $\phi$ production
from $\bar NN$ annihilation}

According to the universal mixing model, Eq.(1), the $\phi/\omega$
production rate from $N\bar N$ annihilation should be less than 1/280, 
and the $\phi\phi/\omega\omega$ rate should be less than $1/80000$.  
The experimental data from LEAR collaborations (cf. [12,25,26]
for full review of the data) show many violations of these predictions.
The JETSET collaboration [27] found 
$\sigma(p\bar p\to\phi\phi) : \sigma(p\bar p\to\omega\omega)\approx 1:150 $
at CM energies around 2.2 GeV, which is more than two orders of magnitude
larger than the prediction of the universal mixing model. The ASTERIX, 
CRYSTAL BARREL and OBELIX collaborations found for $N\bar N$ annihilation
at rest from initial S-wave the following $\phi/\omega$ ratios 
\begin{eqnarray}
B\!R(\phi\gamma) &:& B\!R(\omega\gamma) \approx  1:4 , \\
B\!R(\phi\pi) &:& B\!R(\omega\pi) \approx  1:10 ,\\
B\!R(\phi\omega) &:& B\!R(\omega\omega) \approx  1:50 ,\\
B\!R(\phi\rho) &:& B\!R(\omega\rho) \approx 1:160 ,\\
B\!R(\phi\eta) &:& B\!R(\omega\eta) \approx 1:170 ,\\
B\!R(\phi\pi^+\pi^-) &:& B\!R(\omega\pi^+\pi^-)\approx 1:140 ,\\
B\!R(\phi\pi^0\pi^0) &:& B\!R(\omega\pi^0\pi^0)\approx 1:170 .
\end{eqnarray}
All of them are above the predicted value 1:280, especially, the 
$\phi\gamma$ and $\phi\pi$ channels are more than one order of magnitude
larger than the prediction. In the following, I will examine one by one
the three largest OZI violation channels ($\phi\phi$, $\phi\pi$ and 
$\phi\gamma$) with the hadronic loop mechanism. 
We will see that the $\phi\phi$ and $\phi\pi$ channels can be
explained naturally by hadronic loops while the $\phi\gamma$ channel is
due to the vector meson dominance mechanism. 
For other channels (9-13), no large loop contributions exist [14],
therefore, their ratios are more closer to OZI predictions.

\subsection{$\bar pp\to\phi\phi$}

Its cross section was measured [27] to be $3.7 \mu b$ at the energy of
2.2 GeV while the universal mixing model predicts $0.01 \mu b$.
The universal mixing model is in fact corresponding to a two-loop
diagram shown by Fig.3a. From Sect.2 we know there are
strong cancellation among the $\omega$-$\phi$ mixing loops.
Therefore this kind of diagram should be negligible compared with
one-loop diagrams as shown by Fig.3b,c.

In the unitarity limit (intermediate $K\bar K$ on-shell) for the 
$K\bar K$ loop diagram of Fig.3b, 
all the vertices are well determined by experimental data for the
$\bar pp\to K\bar K$ cross section and $\phi\to K\bar K$ decay width.
The only free parameter is the off-shell cutoff parameter $\Lambda_K$ 
for the t-channel $K$ exchange. With $\Lambda_K=1.2 GeV$, the unitarity
limit for Fig.3b gives a cross section of $2.4 \mu b$ [13]. 
The unitarity limit includes only the imaginary part of the amplitude.
Usually we expect the real part of the amplitude has a similar order of
magnitude to the imaginary part. Then the Fig.3b alone can reproduce the
large measured cross section.
This calculation [13] was criticized [20] for not considering intermediate
$K^*\bar K$, $K\bar K^*$ and $K^*\bar K^*$ states. According to [5],
the $K^*\bar K+K\bar K^*$ loops may have opposite phase to $K\bar K$ and
$K^*\bar K^*$ loops, and therefore there may be cancellations to the
result by considering the $K\bar K$ loop alone. 
For the loops including $K^*$, all the vertices are not well known so
that we cannot calculate them reliably. But a general argument [22]
shows that the summation of four loops will give a similar result
to considering only the $K\bar K$ loop. The key point is given in Table 2.
In allowed partial waves for $\bar pp\to\phi\phi$, only half of them can
go through the $K\bar K$ loop while all of them can go through 
$K^*\bar K$, $K\bar K^*$ and $K^*\bar K^*$ loops. Amplitudes from 
different partial waves cannot cancel each other. In Table 2, $A=0$,
$B$ was calculated by [13], $D$ may have opposite phase to $B$ and may
cancel part of $B$, and $C$ stands alone without cancellations.
It is reasonable to assume $C\approx D$. Then even if $D$ is as large as
$B$ and has opposite phase to $B$,
the summation of all loop contributions will still give a similar 
value to considering only the $K\bar K$ loop.

\begin{center}
Table 2. Hadronic loops for  $\bar pp\to\phi\phi$ 
\vspace{0.5cm}

\begin{tabular}{|l|c|c|}
\hline
Allowed Initial States & $K\bar K$ 
& $K\bar{K^*}$,$K^*\bar K$,$K^*\bar K^*$ \\
\hline
$\matrix {S=0, & L=even, &J=L\cr
S=1, &L=odd, &J=L}$ & forbidden (A) & allowed (C)\\\hline
$\matrix {S=1, &L=odd, &J=L+1\cr 
S=1, &L=odd>1, &J=L-1}$ & allowed (B)& allowed (D)\\\hline
\end{tabular}
\end{center}

For Fig.3c, the vertex of $\bar pp\to\Lambda\Lambda$ can
be determined from experimental data; the vertex of $\phi\Lambda\Lambda$
can be determined by SU(3) arguments. The only free parameter 
is the off-shell cutoff parameter $\Lambda_\Lambda$ 
for the t-channel $\Lambda$ exchange. With $\Lambda_\Lambda=1.5 GeV$,
the hyperon loop diagram gives a cross section about $1.5 \mu b$ [20].
The $\Sigma\Sigma$ loop diagram gives much a smaller contribution [20].

From these results for strange meson loops and hyperon loops, we 
see that the large $\bar pp\to\phi\phi$ cross section can be explained
by the hadronic loop mechanism.

\subsection{$\bar pp\to\phi\pi$}

The hadronic loops with $K^*\bar K$ and $\bar K^*K$ intermediate
states (Fig.4a) were found to give a large enhancement for 
$\bar{p}p\rightarrow \phi\pi^0$ from an initial S state [14-19]. 
The $\rho\rho$ loops (Fig.4b) also give some contribution [14-16]. 
Other loops are much smaller [15]. These hadronic loops can explain
the measured $\phi\pi^0$ branching ratio $(5.5\pm 0.7)\times 10^{-4}$ [28]  
from antiproton annihilation in  liquid hydrogen where S-wave annihilation
dominates. 

However, a question is raised [12] why 
$\bar{p}p\rightarrow \phi\pi^0$ is not 
seen in the annihilation from initial P states where $K^*\bar K$ and 
$\bar K^*K$ have a similar branching ratio as from the initial S state.
This fact is used as evidence for discriminating the rescattering
mechanism and favoring a model assuming existence of strange quarks
in the nucleon [12]. It was suggested [17] that 
possible destructive interference between $l=0$ and $l=2$ of the
intermediate $K^*\bar K$ system may result in a small branching ratio
for $\bar pp\to\phi\pi^0$ from initial P states. This argument is very
shaky since it requires that $l=2$ decay of $\bar pp\to K^*\bar K$ 
happens to be of similar strength with opposite phase to $l=0$ decay. 
Here I give another reason for the suppression of $\phi\pi$ from the 
P state, which is more solid and important.

Both [12] and [17] missed an important fact that for $\bar pp$ annihilation
from P states $K^*\bar K$ can come from $^1P_1$, $^3P_1$ and $^3P_2$ states
with both isospin 0 and 1 while $\phi\pi$ can only come from the $^1P_1$ state
with isospin 1. According to various optical potential models for
protonium annihilation [29,30]
the total decay width for the $I=1$ $^1P_1$ state is only about $1/8$ of the
summation of the total decay width for all possible P states to $K^*\bar K$.
The $K^*\bar K$ decay width may not be directly proportional to
the total decay width for different P states due to some dynamic selection
rule. The $K^*\bar K$ decay width from I=1 $^1P_1$ may be much smaller
than 1/8 of the total $K^*\bar K$ decay width from P states.  It is 
reasonable to expect that $K^*\bar K$ from the $I=1$
$^1P_1$ state is only a very small part of $K^*\bar K$ from all the P
states. Only this small part can contribute to the rescattering mechanism
to the $\phi\pi$ final state. This is contrary to the case for $\bar pp$
annihilation from S states where the allowed partial wave ($I=1$ $^3S_1$)
for $\phi\pi$ is found to be dominant for $K^*\bar K$ [31]. 

There is other experimental evidence 
suggesting that the $I=1$ $^1P_1$ state may have a very small total
decay width. First, the ASTERIX Collaboration found the branching ratios for
$\eta\rho$ and $\eta'\rho$ from P states are much smaller than from
S states [26]. The $\eta\rho$ and $\eta'\rho$ from P states can only come
from the $I=1$ $^1P_1$ state. Second, a recent analysis by the OBELIX 
collaboration [32] shows that the branching ratio of $\omega\pi$ from
$I=1$ $^1P_1$ $\bar pp$ annihilation is also compatible with zero.
So the ratio of
$\phi\pi/\omega\pi$ for P state annihilation may be in fact not suppressed.

In summary, the small branching ratio of $\phi\pi$ from the P state
may be due to the small total decay width of the $I=1$ $^1P_1$ state. 
It is desirable to measure among all $K^*\bar K$ productions
from P states how many percent come from the $I=1$ $^1P_1$ state. 
If this suppression effect were still not enough to  explain the small
$\phi\pi$ branching ratio from the P state, we may consider the effect
proposed by [17] and also a possible cancellation effect between $\rho\rho$
and $K^*\bar K$ loops. Therefore conventional physics can explain
both S wave and P wave annihilation for $\bar pp\to\phi\pi^0$ very well.
The explanation for the large branching ratios of $\bar np\to\phi\pi^+$
and $\bar pn\to\phi\pi^-$ [33]
is straightforward from the charge symmetry argument [16].
 
\subsection{$\bar pp\to\phi\gamma$}

The measured branching ratio for this channel is $(1.7\pm 0.4)\times 10^{-5}$
[28]. The hadronic loop contribution was found [14] to be two order 
of magnitude smaller than this value. Here the vector meson dominance
(VMD) mechanism becomes important [14]. In the VMD mechanism shown by Fig.5,
the branching ratio $BR_{\phi\gamma}$ of $\bar pp\to\phi\gamma$ is related 
to the branching ratio $BR_{\phi\rho}$ of $\bar pp\to\phi\rho$ and
$BR_{\phi\omega}$ of $\bar pp\to\phi\omega$ by the following expression 
[14,34],
\begin{equation}
BR_{\phi\gamma}/P_\gamma = g^2_{\gamma\rho}\cdot \left[ 
BR_{\phi\rho^0}/P_{\rho^0}+\frac{1}{9}BR_{\phi\omega}/P_\omega
+\frac{2}{3} cos\beta \sqrt{BR_{\phi\rho^0}/P_{\rho^0}\cdot
\frac{1}{9}BR_{\phi\omega}/P_\omega} \right] ,
\end{equation}
where the $P_x$ are phase space factors, $\beta$ is the unknown phase
between the amplitudes for the intermediate $\phi\rho$ and $\phi\omega$, 
and $g_{\gamma\rho}$ is the 
$\gamma\rho^0$ coupling constant with $g^2_{\gamma\rho}=3\cdot 10^{-3}$ [34].
Using ASTERIX values of $BR_{\phi\rho^0}=(3.4\pm 1.0)\times 10^{-4}$
and $BR_{\phi\omega}=(5.3\pm 2.2)\times 10^{-4}$ [35], assuming a simple
$k^3_x$ form phase space factors for $P_x$, the VMD mechanism gives
a range of $(0.4\sim 2.7)\times 10^{-5}$ for $BR_{\phi\gamma}$ [14], which
covers well the measured value.

\section{Other possible explanations for the abundant $\phi$ production
from $\bar NN$ annihilation}

There are other very intriguing possibilities which could also explain the
abundant $\phi$ production from $\bar NN$ annihilation. Production of
glueball states [8] could explain the large cross section of 
$\bar pp\to\phi\phi$; coupling to broad $s\bar sn\bar n$ states [9,36]
could explain the large branching ratio for the $\phi\pi$ channel; the 
instanton effects [10] and the presence of substantial $s\bar s$ components 
in the $N/\bar N$ wave function [11,12] could explain several channels.
Though the conventional mechanisms (hadronic loops and VMD) can explain
all channels very well, they cannot exclude these new
possibilities since the relative phases between different mechanisms are
not known. However, the abundant $\phi$ production from $\bar NN$ annihilation
cannot be used as conclusive evidence for these new physics.  

In order to distinguish new physics from the conventional physics,
we need to find where the two mechanisms will give definitely different
predictions. Ref.[12] for the $s\bar s$ mechanism gave many predictions.
Here I will examine whether their predictions for $\bar NN$ annihilation 
can distinguish their mechanism from the conventional mechanisms.

\begin{itemize}

\item Prediction 1: maximum enhancement of $\phi$ production in the initial 
$^3S_1$ state and weaker enhancement in the initial $^1S_0$ state. Among 
measured
channels, see Eqs.(7-13),  (8,11) can only come from the $^3S_1$ state,
(7,9,10) can only come from the $^1S_0$ state, (12,13) can come from both
states. The predicted pattern is not obvious at all.

\item Prediction 2: the $\phi\pi/\omega\pi$ ratio declines for P-state and 
in-flight
annihilation. So does the hadronic loop mechanism as discussed in Sect.3.2.

\item Prediction 3: the branching ratio of $\bar pp\to\pi^0f'_2$ from P-states
is possibly as large as $(1\sim 2)\cdot 10^{-3}$. The $f_2$-$f'_2$ mixing 
mechanism gives $(1\sim 2)\cdot 10^{-4}$. There is no hadronic loop 
calculation for $\bar pp\to\pi^0f'_2$ from P-states yet. Ref.[18] calculated
the $K\bar K^*$, $\bar KK^*$ and $\rho\pi$ loop contributions for 
$\bar pp\to\pi^0f'_2$ from the $^1S_0$ state. The result is smaller than 
the contribution from the $f_2$-$f'_2$ mixing mechanism. 
But they did not include a possible larger loop contribution
from an $\eta$-$a_0(980)$ intermediate state.
For $\bar pp\to\pi^0f'_2$, $^1S_0$ and $^3P_{1,2}$ $\bar pp$ states contribute,
while for $\bar pp\to\pi^0\phi$ $^3S_1$ and $^1P_1$ $\bar pp$ states 
contribute. As discussed in Sect.3.2, $K^*\bar K$ are dominantly coming
from $^3S_1$ and $^3P_{1,2}$ states; we expect $K^*\bar K+K\bar K^*$ loops 
give a very large contribution to $\bar pp\to\pi^0f'_2$ from P-states and
$\bar pp\to\pi^0\phi$ from the S-state, but small contributions to
$\bar pp\to\pi^0f'_2$ from the S-state and $\bar pp\to\pi^0\phi$ from P-states.
Once again the hadronic loop mechanism predicts a similar thing here as the
$s\bar s$ mechanism.

\item Prediction 4: $\bar pp\to\phi\phi$ is more enhanced for initial 
spin-triplet states. Here the $K\bar K$ loop is allowed for initial 
spin-triplet states, but is forbidden from initial spin-singlet states.

\item Prediction 5: in $\bar pp\to\phi\pi^+\pi^-$, $^3S_1$ should dominate.
ASTERIX [35] found $^1S_0$ dominates while OBELIX [37] found $^3S_1$ dominates. 
These contradictory experimental results remain to be clarified.
Even if the OBELIX result is correct, it is still not contradictory to
conventional physics. The $^3S_1$ has a larger statistical weight than
the $^1S_0$ state, i.e., 3:1. 
The branching ratio for this channel is not
far away from the OZI prediction, see Eq.(12). Several hadronic loops
can have small contributions. They may or may not cancel this statistical
weight effect. 

\item Prediction 6: $\phi\gamma/\omega\gamma$ should increase for P-state
annihilation. This can also explained by the VMD mechanism since the 
$\rho\phi$ and $\omega\phi$ branching ratios were found to be increasing for
P-state annihilation [35].

\item Prediction 7: $\bar pp\to\phi\pi$, $\omega\pi$ should have different 
angular distributions due to different production mechanisms. In the
hadronic loop mechanism, $\bar pp\to\phi\pi$, $\omega\pi$ also have
different production mechanisms. The calculated angular distribution
for $\bar pp\to\phi\pi$ in the hadronic mechanism was found to be
compatible with experimental data [19].

\item Prediction 8: large $\phi/\omega$ ratio in the Pontecorvo reaction
$\bar pd\to\phi n$.
This is also consistent with the hadronic loop mechanism. Since 
the reaction can go through both $\phi\pi^0n$ and $\phi\pi^-p$ intermediate
states, the interference effect can make the $\phi/\omega$ ratio here
larger than for $\bar pN\to\phi\pi$ as reported by the 
OBELIX collaboration [38].
\end{itemize}

After the above detailed examination, no clear-cut prediction for $\bar NN$ 
annihilation is found to distinguish the new $s\bar s$ mechanism.   
We note in passing that evidence for the presence of $s\bar s$ in the nucleon
from other sources [11] was criticized by [39,40].

\section{Conclusion}

\noindent 1) Hadronic loops can explain all $s\bar s$-$n\bar n$ mixing 
naturally.
$1^{--}$, $2^{++}$ and $3^{--}$ nonets have smaller $s\bar s$-$n\bar n$
mixing due to strong cancellations between $K\bar K^*+\bar KK^*$ loops
and $K\bar K+K^*\bar K^*$ loops; $0^{-+}$, $0^{++}$, $1^{+-}$ and $1^{++}$
nonets have larger mixing due to selection rules against either $K\bar K$ or
$K\bar K+K^*\bar K^*$ loops, which leads to weaker or no cancellation.

\noindent 2) No nearly pure $s\bar s$ $0^{++}$ meson exists due to a large 
$K\bar K$ loop without centrifugal barrier factor and cancellation from
$\bar KK^*+K\bar K^*$ loops.
  
\noindent 3) Conventional mechanisms (hadronic loops and vector meson
dominance) can explain all $\phi$ enhancements from $\bar NN$ annihilation
naturally.

\noindent 4) No conclusive evidence from $N\bar N$ annihilation exists for 
glueball states,
$s\bar sn\bar n$ four quark states, instanton effects, and strange quarks 
in the nucleon, though they are not excluded.

\noindent 5) There is no simple clear-cut prediction for $\bar NN$ 
annihilation to prove the presence of $s\bar s$ in nucleon yet. 
Not only $\phi$, $f'_2$ production channels but also related channels
($\bar KK$, $K\bar K^*$, $\bar K^*K^*$, $\rho\rho$, $a_0\eta$, etc.)
should be investigated with detailed partial wave analyses to see
whether we can find a place where the hadronic loops fail.
Only after such hard detailed studies, may we claim any conclusive
evidence for new physics.

\bigskip\bigskip
\noindent {\bf Acknowledgement:} I am very grateful to the organizers of 
NAN95 for invitation to give the talk and warm hospitality in Moscow. I thank
V.V.Anisovich, C.J.Batty, D.V.Bugg, F.E.Close, H.J.Lipkin, M.P. Locher, 
N.I.Kochelev, V.E. Markushin, M.G. Ryskin, M.G. Sapozinikov, A.V.Sarantsev, 
I.Scott, N.T\"ornqvist for helpful discussions and comments.

\newpage
\centerline{\bf Figure Captions}

\noindent Fig.1. OZI (a) allowed, (b) forbidden diagrams.

\noindent Fig.2. Diagrams for $s\bar s\leftrightarrow n\bar n$ transition:
(a) first order; (b) hadronic loops.

\noindent Fig.3. Loop diagrams for $\bar pp\to\phi\phi$: (a) $\omega$-$\phi$
mixing mechanism;  (b) $K\bar K$ meson loop; (c) $\Lambda\bar\Lambda$ 
hyperon loop.

\noindent Fig.4. (a) $\bar KK$ and (b) $\rho\rho$ meson loops for 
$\bar pp\to\phi\pi^0$.

\noindent Fig.5. Vector meson dominance mechanism for $\bar pp\to\phi\gamma$
through (a) $\rho\phi$ and (b) $\omega\phi$ intermediate states.

\newpage
\begin{figure}[htbp]
\begin{center}\hspace*{-0.cm}
\epsfysize=15.0cm
\epsffile{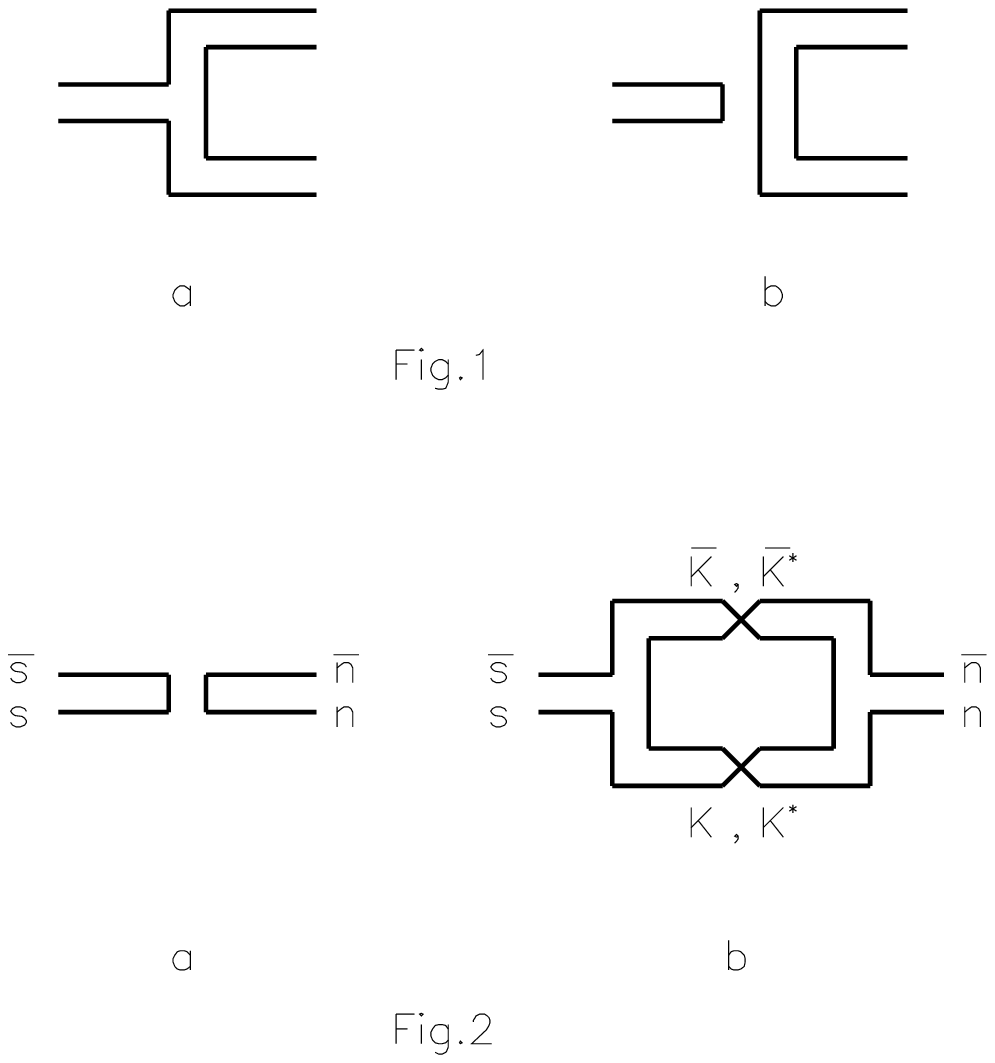}
\end{center}
\caption{
}
\label{fig:fig1}
\end{figure}
\begin{figure}[htbp]
\begin{center}\hspace*{-0.cm}
\epsfysize=1.0cm
\end{center}
\caption{
}
\label{fig:fig2}
\end{figure}

\newpage
\begin{figure}[htbp]
\begin{center}\hspace*{-0.cm}
\epsfysize=18.0cm
\epsffile{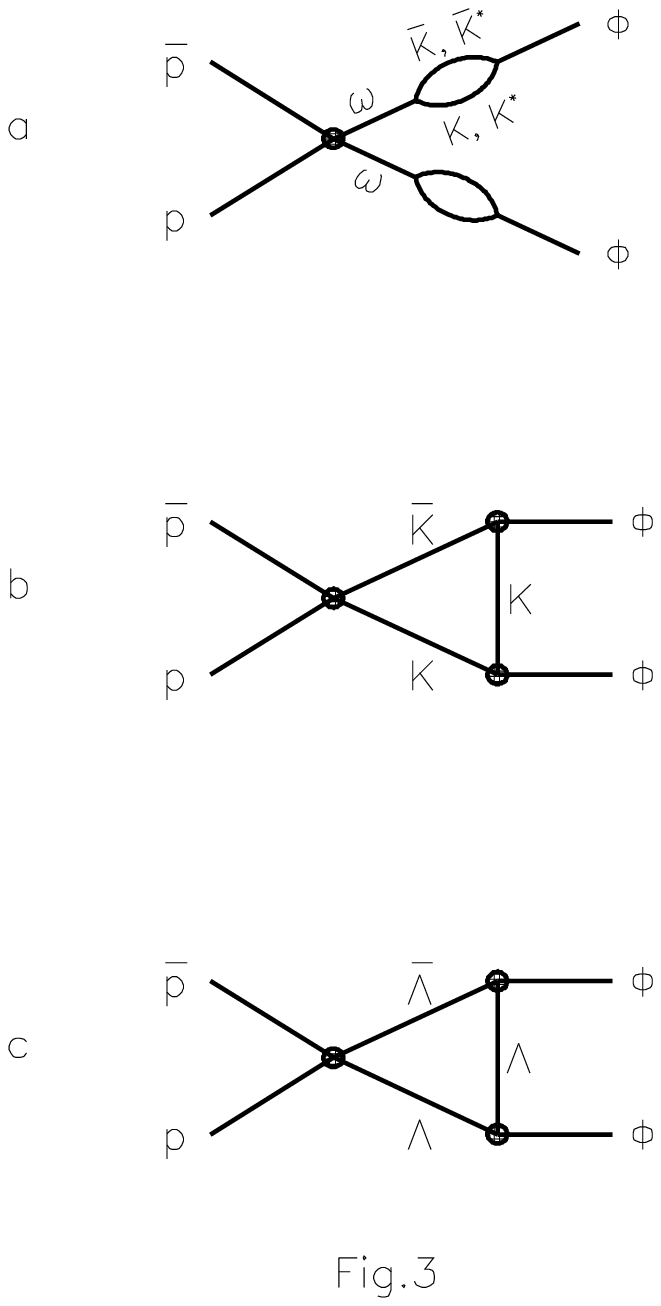}
\end{center}
\caption{
}
\label{fig:fig3}
\end{figure}

\newpage
\begin{figure}[htbp]
\begin{center}\hspace*{-0.cm}
\epsfysize=15.0cm
\epsffile{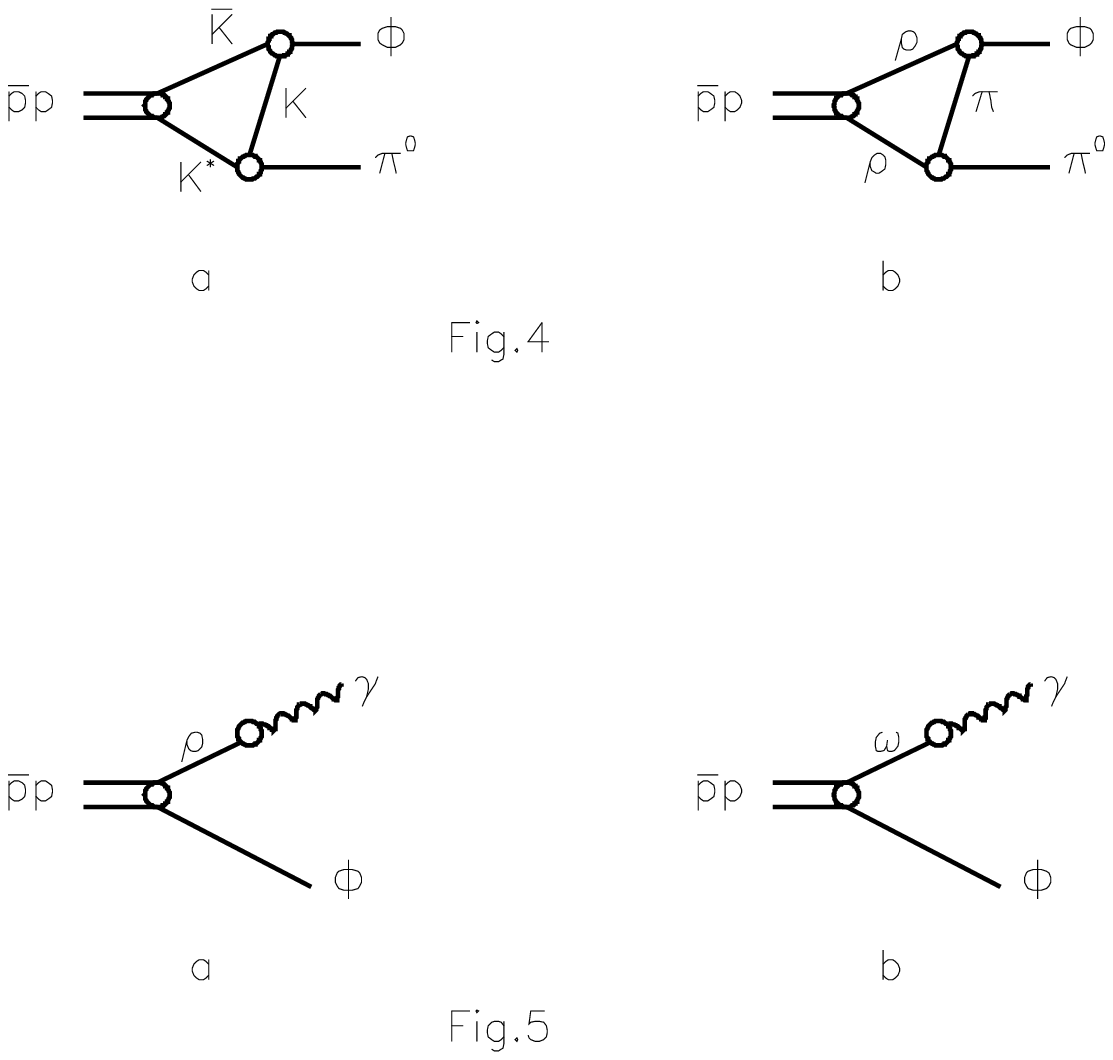}
\end{center}
\caption{
}
\label{fig:fig4}
\end{figure}
\begin{figure}[htbp]
\begin{center}\hspace*{-0.cm}
\epsfysize=1.0cm
\end{center}
\caption{
}
\label{fig:fig5}
\end{figure}
\end{document}